\documentclass[letter,longauth]{aa}
\usepackage{natbib}
\bibpunct{(}{)}{;}{a}{}{,} % to follow the A&A style
\usepackage{booktabs}
\usepackage{verbatim,multirow,multicol}
\usepackage{subfig}
\usepackage{hhline}
\usepackage{soul}
\setstcolor{red}
\usepackage{color}
\usepackage{tablefootnote}

\definecolor{Red}{rgb}{0.9,0,0}
\definecolor{Blue}{rgb}{0,0,0.9}
\definecolor{Green}{rgb}{0,0.5,0}
\definecolor{Black}{rgb}{0,0,0}

\newcommand{\Gaia}{\textit{Gaia}~}

\usepackage{txfonts}
\usepackage{graphicx}
\usepackage{multirow}
\usepackage{booktabs}
\usepackage{hyperref}

\begin{document} 

 \title{The changing material around (2060) Chiron from an occultation on 2022 December 15}

 % \title{A stellar occultation by (2060) Chiron on 2022 December 15$^{\rm th}$ while still in brightness outburst}

% \subtitle{Chiron extinction features from an occultation}

%J. L. Ortiz, C. L. Pereira, B. Sicardy, F. Braga-Ribas, A. Takey, A. Fouad, A. Shaker, S. Kaspi, N. Brosch, M. Kretlow, R. Leiva, B. Morgado, N. Morales, M. Vara-Lubiano, P. Santos-Sanz, E. Fernandez-Valenzuela, R. Duffard, F. L. Rommel, Y. Kilic, O. Erece, D. Koseoglu, E. Ege, J. Desmars, R. Morales, A. Alvarez-Candal, J. L. Rizos, J. M. Gómez-Limón, R. Sfair, T. de Santana, M. Assafin, R. Vieira-Martins, A. R. Gomes-Júnior, J. I. B. Camargo, J. Lecacheux, anybody else??

  \author{J.~L.~Ortiz\inst{1} % email: ortiz@iaa.es,orcid: 0000-0002-8690-2413
  \and
  C.~L.~Pereira\inst{2,3} % email: chrystianpereira@on.br,orcid: 0000-0003-1000-8113
  \and
  B.~Sicardy\inst{4} % email: bruno.sicardy@obspm.fr,orcid: 0000-0003-1995-0842
  \and
  F.~Braga-Ribas\inst{5,2,3} % email: felipebribas@gmail.com,orcid: 0000-0003-2311-2438
  \and
  A.~Takey\inst{6} % 
  \and
  A.~M.~Fouad\inst{6} % 
  \and
  A.~A.~Shaker\inst{6} % 
  \and
  S.~Kaspi\inst{7} % 
  \and
  N.~Brosch\inst{7} % 
  \and
  M.~Kretlow\inst{1,8} % email: mike@kretlow.de,orcid: 0000-0001-8858-3420
  \and
  R.~Leiva\inst{1} %
  \and
  J.~Desmars\inst{9,10} % email: josselin.desmars@obspm.fr,orcid: 0000-0002-2193-8204
  \and
  B.~E.~Morgado\inst{11,2,3} % email: morgado.fis@gmail.com,orcid: 0000-0003-0088-1808
  \and
  N.~Morales\inst{1} % 
  \and
  M.~Vara-Lubiano\inst{1} % email: mvara@iaa.es,orcid: 0000-0002-8112-0770
  \and
  P.~Santos-Sanz\inst{1} % email: psantos@iaa.es,orcid: 0000-0002-1123-9830
  \and
  E.~Fernández-Valenzuela\inst{12,1} % email: estela@ucf.edu,orcid: 0000-0003-2132-7769
  \and
  D.~Souami\inst{4,13,14}
  \and
  R.~Duffard\inst{1} % 
  \and
  F.~L.~Rommel\inst{5,2,3} % 
  \and
  Y.~Kilic\inst{15,16} % email: yucelkilic1@gmail.com,orcid: 0000-0001-8641-0796
  \and
  O.~Erece\inst{16} % 
  \and
  D.~Koseoglu\inst{16} % 
  \and
   E.~Ege\inst{17} % 
  \and
  R.~Morales\inst{1} % 
  \and
  A.~Alvarez-Candal\inst{1,18} % 
  \and
  J.~L.~Rizos\inst{1} % 
  \and
  J.~M.~Gómez-Limón\inst{1} % 
  \and
 % R.~Sfair\inst{18,19} % 
 % \and
 % T.~de~Santana\inst{4,19} % 
 % \and
  M.~Assafin\inst{11,3} % email: massaf@ov.ufrj.br,orcid: 0000-0002-8211-0777
  \and
  R.~Vieira-Martins\inst{2,3} % email: rvm@on.br,orcid: 0000-0003-1690-5704
  \and
  A.~R.~Gomes-Júnior\inst{19,20,3} % email: altairgomesjr@gmail.com,orcid: 0000-0002-3362-2127
  \and
  J.~I.~B.~Camargo\inst{2,3} % email: camargo@on.br,
  \and
  J.~Lecacheux\inst{4} % 
  }

  \institute{Instituto de Astrofísica de Andalucía ? Consejo Superior de Investigaciones Científicas, Glorieta de la Astronomía S/N, E-18008, Granada, Spain\\
  \email{ortiz@iaa.es}
  \and
  Observat\'orio Nacional/MCTI, R. General Jos\'e Cristino 77, CEP 20921-400 Rio de Janeiro - RJ, Brazil
  \and
  Laborat\'orio Interinstitucional de e-Astronomia - LIneA, Rio de Janeiro, RJ, Brazil
  \and
  LESIA, Observatoire de Paris, Universit\'e PSL, Sorbonne Universit\'e, Universit'e de Paris, CNRS, 92190 Meudon, France
  \and
  Federal University of Technology - Paran\'a (UTFPR-Curitiba), Rua Sete de Setembro, 3165, CEP 80230-901, Curitiba, PR, Brazil
  \and
  National Research Institute of Astronomy and Geophysics (NRIAG), Helwan 11421, Cairo, Egypt
  \and
  The Wise Observatory and the Raymond and Beverly Sackler School of Physics and Astronomy, The Faculty of Exact Sciences, Tel Aviv University, Tel Aviv 69978, Israel
  \and
 % Internationale Amateursternwarte (IAS) e. V., Mittelstr. 6, 15749 Mittenwalde, Germany
 % \and
  International Occultation Timing Association - European Section (IOTA/ES), Am Brombeerhag 13, 30459 Hannover, Germany
  \and
   Institut Polytechnique des Sciences Avanc\'ees IPSA, 63 boulevard de Brandebourg, 94200 Ivry-sur-Seine, France
  \and
   Institut de Mécanique C\'eleste et de Calcul des \'Eph\'em\'erides, IMCCE, Observatoire de Paris, PSL Research University, CNRS, Sorbonne Universit\'es, UPMC Univ Paris 06, Univ. Lille, France
  \and
  Universidade Federal do Rio de Janeiro - Observat\'orio do Valongo, Ladeira Pedro Antônio 43, CEP 20.080-090, Rio de Janeiro - RJ, Brazil
  \and
  Florida Space Institute, UCF, 12354 Research Parkway, Partnership 1 Building, Room 211, Orlando, USA
  \and
Departments of Astronomy, and of Earth and Planetary Science, 501 Campbell Hall University of California, Berkeley, CA 94720, United States of America
  \and
  naXys, University of Namur, 8 Rempart de la Vierge, Namur, B-5000, Belgium
  \and
  Akdeniz University, Faculty of Sciences, Department of Space Sciences and Technologies, 07058 Antalya, Turkey
  \and
  TÜB\.{I}TAK National Observatory, Akdeniz University Campus, 07058 Antalya, Turkey
  \and
  Istanbul University, Faculty of Science, Department of Astronomy and Space Sciences, 34116, Beyaz?t, Istanbul, Turkey
  \and
  Instituto de F\'isica Aplicada a las Ciencias y las Tecnolog\'ias, Universidad de Alicante, San Vicent del Raspeig, E03080, Alicante, Spain
  \and
  Institute for Astronomy and Astrophysics, Eberhard Karls University of Tübingen, Tübingen, Germany
  \and
  Orbital Dynamics and Planetology Group, UNESP - São Paulo State University, Guaratinguetá, Brazil
% \and
% Institute of Physics, Federal University of Uberlândia, Uberlândia-MG, Brazil
  }

 % \institute{Instituto de Astrofisica de Andalucia,
 % Glorieta de la astronomia sn, 18008 Granada, Spain\\
 % \email{ortiz@iaa.es}
 % \and
 % Observat\'orio Nacional/MCTIC, R. General Jos\'e Cristino 77, Bairro Imperial de S\~ao Crist\'ov\~ao, Rio de Janeiro (RJ),
 % Brazil\\

%A. Takey, A. M. Fouad, A. A. Shaker is:
%National Research Institute of Astronomy and Geophysics (NRIAG),
%11421, Cairo, Helwan, Egypt

%S. Kaspi, N. Brosch
%The Wise Observatory and the Raymond and Beverly Sackler School of
%Physics and Astronomy, The Faculty of Exact Sciences, Tel Aviv
%University, Tel Aviv 69978, Israel

      % \email{c.ptolemy@hipparch.uheaven.space}
      % \thanks{The university of heaven temporarily does not accept e-mails}
 % \and
 % University of Alexandria, Department of Geography, Egypt\\
 % }

  \date{Received March 15, 2023; accepted }

\abstract{We could accurately predict the shadow path and successfully observe an occultation of a bright star by Chiron on 2022 December 15. The Kottamia Astronomical Observatory in Egypt did not detect the occultation by the solid body, but we detected three extinction features in the light curve that had symmetrical counterparts with respect to the central time of the occultation. One of the features is broad and shallow, whereas the other two features are sharper with a maximum extinction of $\sim$25$\%$ at the achieved spatial resolution of 19 km per data point. From the Wise observatory in Israel, we detected the occultation caused by the main body and several extinction features surrounding the body. When all the secondary features are plotted in the sky plane 
%we find that they can be caused by a double ring structure with radii of 325 $\pm$ 16 km and 423 $\pm$ 11 km embedded in a broad disk of $\sim$590 km in diameter that surrounds Chiron. The disk would have at least two dust or ice concentrations at the ring radii. At least one of these structures appears to be outside the Roche limit. 
we find that they can be caused by a broad $\sim$580 km disk with
concentrations at radii of 325 ± 16 km and 423 ± 11 km surrounding
Chiron. At least one of these structures appears to be outside the
Roche limit.
The ecliptic coordinates of the pole of the disk are $\lambda$ = 151$^\circ~\pm$ 8$^\circ$ and $\beta$ = 18$^\circ~\pm$ 11$^\circ$, in agreement with previous results. We also show our long-term photometry indicating that Chiron had suffered a brightness outburst of at least 0.6 mag between March and September 2021 and that Chiron was still somewhat brighter at the occultation date than at its nominal pre-outburst phase. The outermost extinction features might be consistent with a bound or temporarily bound structure associated with the brightness increase. However, the nature of the brightness outburst is unclear, and it is also unclear whether the dust or ice released in the outburst could be feeding a putative ring structure or if it emanated from it.}


  \maketitle
%
%-------------------------------------------------------------------

\section{Introduction}

Soon after the unexpected discovery of a dense double ring separated by a small gap in the Centaur (10199) Chariklo from a well-observed stellar occultation \citep{BragaRibas2014}, a natural question was whether that ring structure was something unique to Chariklo or whether there could be similar structures in other Centaurs or related bodies in the outer solar system. In that context, the claim that the Centaur (2060) Chiron might also have a ring system \citep{Ortiz2015} based on a few observed stellar occultations reported in the literature \citep{Elliot1995,Ruprecht2015,Bus1996,Sickafoose2020}, in combination with photometric and spectroscopic hints, was not as fully convincing as the Chariklo case, mainly because Chariklo's observations were obtained from several locations, not just two sites. Particularly intriguing was that some extinction features were not seen at all longitudes around Chiron, meaning that the putative ring system would be incomplete or highly inhomogeneous and quite different from Chariklo's ring system. Besides, there were previous claims that extinction events on an occultation by Chiron might be due to cometary-like jets \citep{Elliot1995}.

Later, another ring system was unambiguously detected, this time around the large trans-Neptunian object (TNO) and dwarf planet Haumea \citep{Ortiz2017}. In addition, yet another ring system has just been announced around the large TNO Quaoar \citep{Morgado2023,Pereira2023}. This time the ring is outside Quaoar's Roche limit, with at least a very dense arc and tenuous material coexisting in the ring structure, meaning that such a ring is highly inhomogeneous in longitude. In this new context, the inhomogeneity of the potential ring around Chiron would not be that surprising.

Predicting and observing new stellar occultations was very relevant to characterize the material around Chiron properly. Within the Lucky Star collaboration \footnote{\url{https://lesia.obspm.fr/lucky-star}} a promising event was identified for 2022 December 15, successfully observed, and preliminary results are reported here.

\section{Observations of the occultation}

Observations were obtained with the 1.88m telescope at Kottamia Astronomical Observatory (KAO) in Egypt \citep{Azzam2010}, equipped with the Kottamia Faint Imaging Spectro-Polarimeter, KFISP \citep{Azzam2022}. %(reference Azzam et al. 2022 https://ui.adsabs.harvard.edu/abs/2022ExA....53...45A/abstract). 
We used the imaging mode in binning 4x4 to decrease the readout time as much as possible. The Sloan Digital Sky Survey SDSS g-band filter was used. Integration time was 3s, and the average readout time was around 1.5s, meaning that the full cycle of consecutive observations typically took 4.5s. Given that the speed of Chiron with respect to the observer was $4.24\,\rm{km\,s}^{-1}$, the achieved spatial resolution was 19 km. The observing sequence started at 16:56:09.7 UT and finished at 18:10:51 UT. The sky was clear. 

The Wise Observatory (Israel) observations consisted of images starting at 17:25:49 and finishing at 17:50:52 UT. The start of image acquisition was a bit later than planned, which resulted in missing part of the phenomena, as shown in sections 2 and 3. The observations were taken with a 0.45m telescope equipped with a thermoelectrically cooled CCD camera based on a Kodak KAF-8300 chip. Exposure time was 3s with a readout time of approximately 4.5s. The full observing cycle was thus around 7.5s, and the achieved spatial resolution was 31.8 km. No filters were used. The sky was clear.

Other observations were attempted from another telescope at Wise Observatory and Neot Smadar Observatory in Israel, but technical and operational problems prevented observations at the occultation time. At T\"UBITAK National Observatory in Turkey, observations were obtained with the 1-meter telescope. 
%(give more details here regarding instrument etc). 
The sky was cloudy with intermittent gaps that allowed some observations at the predicted time of the occultation, but no occultation was detected. The star coordinates and other relevant data are shown in Table 1. A table summarizing the observing circumstances at the involved sites is given in Table 1 of the Appendix.

\begin{table}[!ht]
  \setlength{\tabcolsep}{2mm} 
  \centering
  \caption{
  Occulted star information. 
  }
  \begin{tabular}{l l}
  \toprule \toprule
  Epoch & 2022-12-15 17:36:52 UTC \\
  Source ID & \Gaia DR3 2556932574069626624 \\
  Star position & $\alpha_\star$= 00$^h$39$^m$58$^s$.71911 $\pm$ 0.1233 mas \\
  \hspace{5mm}at epoch\tablefootmark{1} & $\delta_\star$= 06$^\circ$16$'$08$''$.8589 $\pm$ 0.1451 mas \\
  Magnitudes\tablefootmark{2} & G = 12.7; J = 12.1; H = 12.0; K = 12.0 \\
  Apparent diameter & 0.02 mas / 0.21 km \\
  \midrule
  \end{tabular}
  \tablefoot{
  \tablefoottext{1}{The star position was taken from the \Gaia Data Release 3 (GDR3) star catalog \citep{GaiaColab2022} and is propagated to the event epoch using the formalism of \citet{Butkevich2014} applied with SORA \citep{GomesJr2022}. The duplicated source flag in GDR3 is 1, meaning that the star might have had some processing errors or be multiple, but the flux of the star went to zero at the occultation by the main body, meaning that the star is not multiple or the potential companion would be very dim.
  \tablefoottext{2}{J, H and K from the 2MASS catalog.}
  }
  }
  \label{tab:star_obj}
\end{table}

\section {Data reduction and analysis}

% The data were managed through the Occultation portal website. \footnote{http://occultation.tug.tubitak.gov.tr}
The images, compiled and managed through the Tubitak Occultation Portal website \citep{Kilic2022}, were bias subtracted and flatfielded. Median bias frames and median sky flat frames were used.
Aperture photometry of the target star blended with Chiron from the image sequences was carried out using different synthetic apertures to get the least dispersion possible in the relative photometry. Comparison stars of similar brightness were used to derive the relative photometry to compensate for small atmospheric fluctuations. The methods used were the same as those described in 
\cite{Ortiz2020b}.
%(Ortiz et al. 2020). 
The best light curves in terms of dispersion are shown in Figs. 1 and 2, where the light curves have been normalized to 1 outside of the occultation time.

The light curves show clear brightness drops that do not reach zero flux in the case of Kottamia observations, meaning that the solid body of Chiron did not occult the star as observed from Kottamia, but that there was material around the body causing the extinction of the light to different degrees up to $\sim$25\% of the flux. The timings of the features in the light curve appear symmetrical to the time of the closest approach of the body to the star. The symmetrically located extinction features may not reach the same depth because they might have happened closer or farther to the moments of readout in the observing sequence, causing apparent variability in the extinction level. However, it is also possible that the particles causing the extinction are spread in an area of different widths, or the optical properties of the particles may be slightly different at the symmetric locations.

The light curve at Wise Observatory shows that the main body was occulted because the flux dropped to zero, but secondary features are also seen in the light curve. Unfortunately, the data acquisition did not start soon enough to record the counterparts of all the extinction features seen in the second part of the light curve after the occultation of the main body.

%From the light curves, we extracted the times at which the sharp drops

\begin{figure}
\centering
    %\includegraphics[width=\columnwidth]{experimentoFiguraChironV4.png}
    \includegraphics[width=\columnwidth]{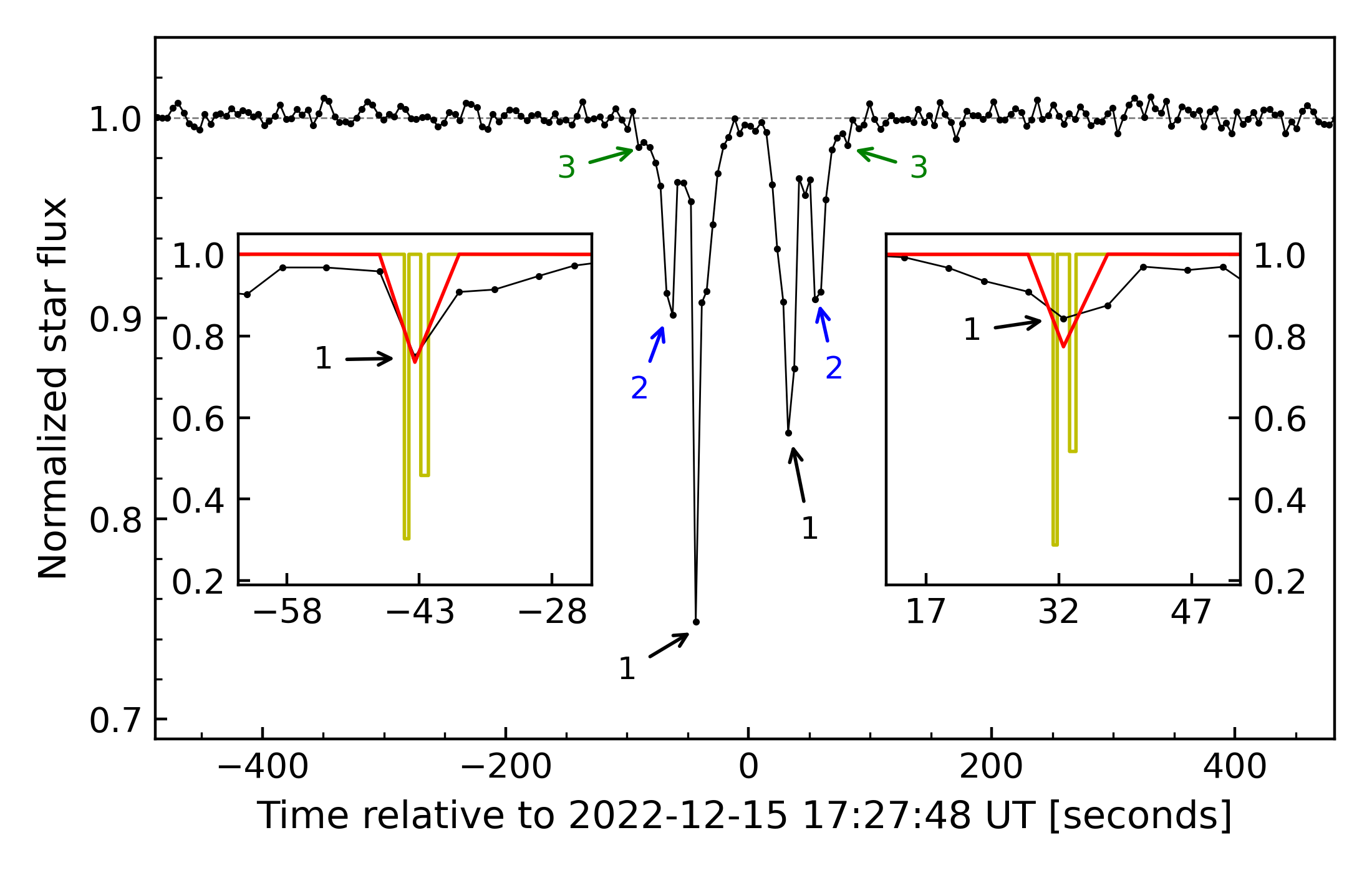}
  \caption{Occultation light curve from Kottamia observatory. Flux versus time is shown in black, and the dotted grey line corresponds to a linear fit to the baseline of the light curve. The main symmetric extinction events are indicated with arrows and numbered. The inserts show magnified views around the deepest extinction events (black line) together with a model of the ring structure compatible with the 2011 occultation (gold line), and the model convolved to the time resolution achieved in 2022 (red line). The red line reproduces the maximum drops but not the rest of the curve, indicating more material around Chiron in 2022 than in 2011. See text.}
  \label{Kottamia}
\end{figure}

\begin{figure}
\centering
    %\includegraphics[width=\columnwidth]{experimentoFigura2ChironV5.png}
    \includegraphics[width=\columnwidth]{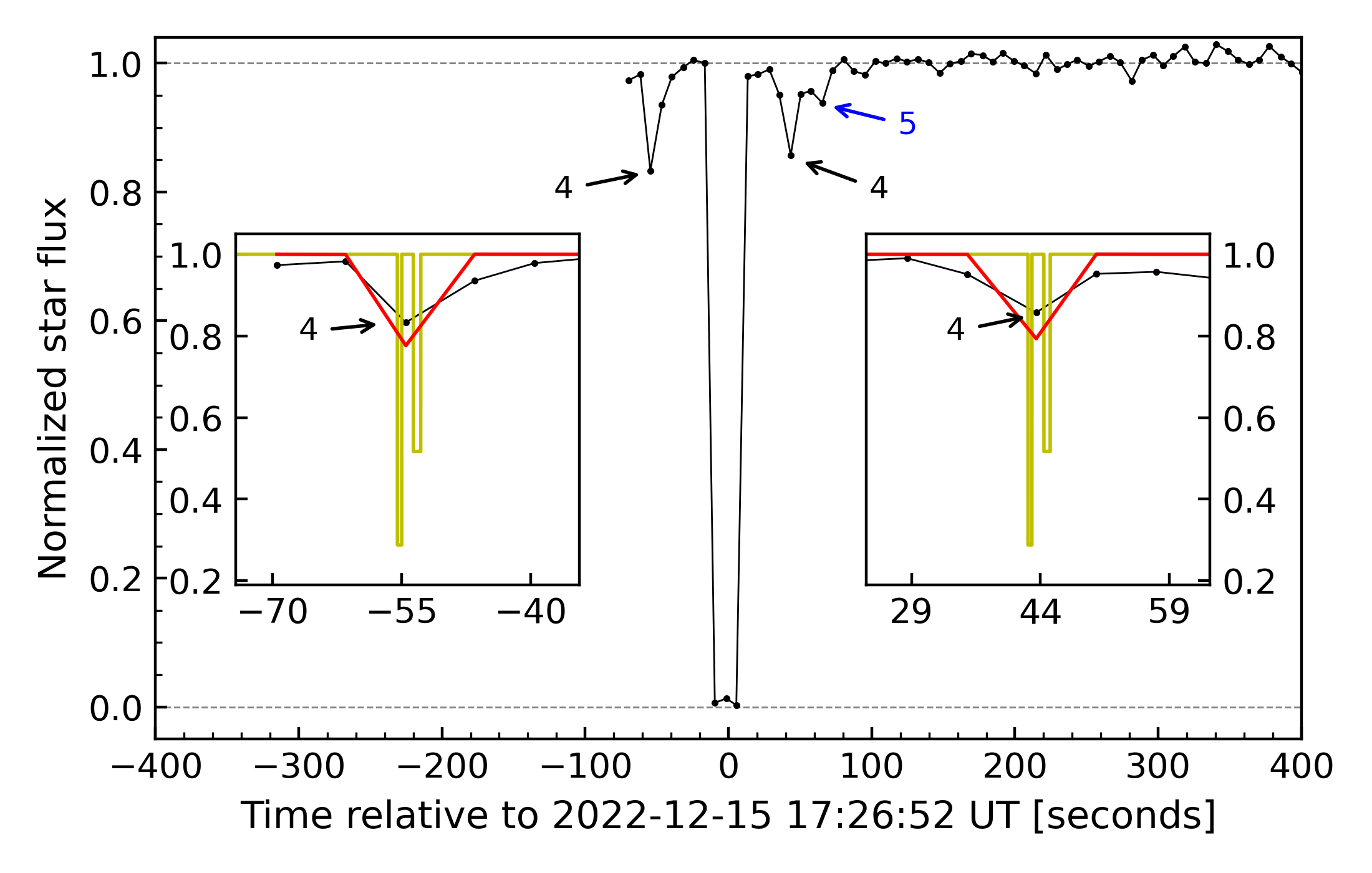}
  \caption{
  %\st{Occultation light curve from Wise observatory. Flux versus time is shown. The flux drops to zero near minute 1046.7, meaning that the main body occulted the star, but there were also extinction events indicated with numeric labels.} 
  Same as Fig. \ref{Kottamia}, but for Wise Observatory. The central drop to zero flux is due to the main body occultation. }
  \label{Wise}
\end{figure}

\section{Interpretation}

\begin{figure}
\centering
    \includegraphics[width=\columnwidth]{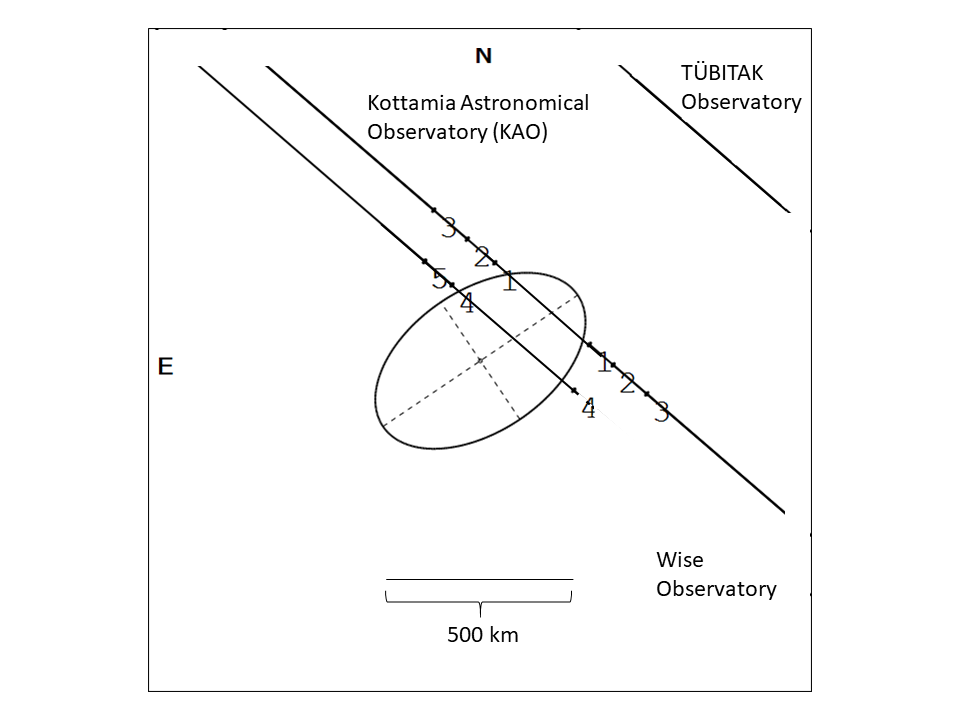}

  \caption{The black straight lines show the chords in the plane of the sky corresponding to the three sites that obtained observations, with the moments corresponding to the extinction events indicated with the same numeric labels as in the light curves. The segment for Wise observatory is shorter than that of Kottamia because image acquisition started later than planned. The black ellipse corresponds to the nominal ring proposed in \cite{Ortiz2015}, which requires some changes to be consistent with the extinction features 1 and 4. The dotted lines show the axes of the ellipse. The motion of the star is from right to left. A better model for the ring is shown in Fig. \ref{proposedrings}.}
  \label{anillodepartida}
\end{figure}

The moments of the extinction features shown in the light curves and identified with labels 1 to 5 are plotted in the sky plane in Fig.~\ref{anillodepartida}. 
%The chord of the central body is also depicted.
In \cite{Ortiz2015}, it was proposed that Chiron could have a ring with a radius around 324 km and pole ecliptic coordinates of $\lambda = 144^\circ$ and $\beta = 24^\circ$ based on the simultaneous analysis of previous occultation observations and long term photometry, with an uncertainty of around 10$^\circ$.
\cite{Sickafoose2020} presented an in-depth analysis of occultation extinction features in the 2011 occultation event that they recorded and concluded that the proposed ring seemed consistent with their observations. Using the pole above, the current position angle of the minor axis of the ring ellipse should be 34.7$^\circ$, and the aspect angle of the ring should be 52.9$^\circ$. Because the proposed ring had a radius 324~km, we can draw the expected elliptical configuration and see how close this ring system is to the deepest secondary occultation symmetric features labeled with numbers 1 and 4. The center of the ring was chosen as the center of the solid body because the ring is expected to be equatorial. For the solid body, we assumed an elliptical shape concentric with the rings because the body is expected to be non-spherical since it shows a shape-dominated rotational light curve with a period of 5.917813 h \citep{Marcialis1993}. By looking at the drawn ellipse, there is reasonable agreement because the ring is close to the extinction features, but a better match is obtained by keeping the position angle at 34 degrees and allowing for a smaller aspect angle, around 45$^\circ$. The best visual match is achieved by allowing for a Position Angle of 45.3$^\circ$ and an aspect angle of 45.6$^\circ$ for the ring ellipse. This requires pole coordinates $\lambda=149^\circ$ and $\beta=14^\circ$, a solution within the 10$^\circ$ uncertainty quoted in \cite{Ortiz2015}.

\begin{figure}
\centering
    \includegraphics[width=\columnwidth]{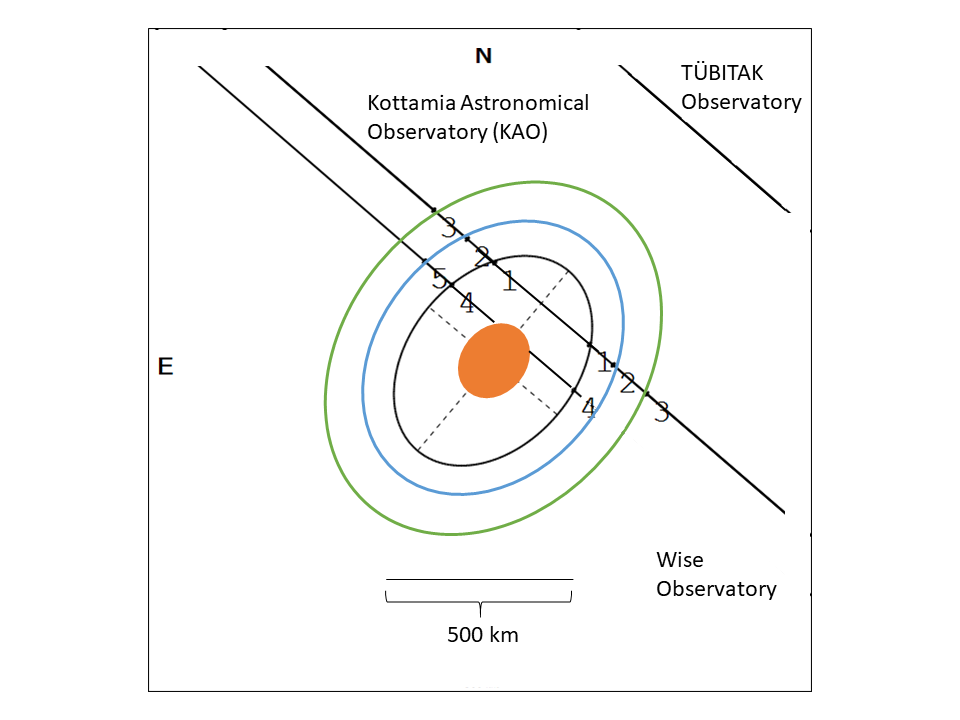}

  \caption{Potential ring structure projected in the plane of the sky that can explain the different extinction features observed in the occultation light curves. The ellipses correspond to the rings that best match the extinction features. The central filled ellipse in brown corresponds to a possible projection of the nucleus compatible with the main body occultation detected from the Wise Observatory. The motion of the star is from right to left.}
  \label{proposedrings}
\end{figure}

A concentric structure to this ellipse can fit the positions of the other extinction features seen at both Kottamia and Wise, as %\st{. This is} 
depicted as a blue ellipse in Fig. \ref{proposedrings}. The radius of this additional potential ring is $\sim$423 km. The broad and faint extinction features only seen at Kottamia can be consistent with a broad ring or disk that extends to $\sim$ 590 km away from the center of the body. Fig.\ref{proposedrings} depicts this with a green ellipse.

%In figure XXX we show a sketch indicating the phenomenology seen.

%More accurate values for the ring diameters and ring pole can be obtained by simultaneously fitting the extinction features observed in this occultation together with the extinction features reported by Sickafoose et al. (2019) for the occultation in 2011, which are the two best-observed occultations. 
%When this is done, the refined pole coordinates for the rings become XX and XX. 

\subsection{Joint interpretation of the occultation features in 2022 and the occultation in 2011}

%{\bf Include here all the stuff from Chrystian}

% \blu 

  The stellar occultation by Chiron observed on November 29, 2011, UT \citep{Ruprecht2015, Sickafoose2020} allowed symmetrical secondary structures to be detected around this Centaur. Using the time of each extinction feature presented in those works, we can put the 2011 Faulkes North Telescope (FTN) and NASA Infrared Telescope Facility (IRTF) detections onto the sky plane, considering the necessary projections.

  The first step is to fit the occultation by the main body. The simplest model with only one positive chord is the circular one. Thus, we fit a circle to the extremities of the occultations chords using the equivalent diameter of 210 km \citep{Lellouch2017}. Although we know that Chiron does not have a spherical shape, our fits consider a radial uncertainty of 10 km for the spherical model. Thus, a possible displacement of the body's center due to its triaxial shape would keep our results reliable. 

  Using the pole orientation derived by \cite{Ortiz2015}, we can analyze the distribution of detections in the sky plane and how they line up with ring-like structures around Chiron. At first glance, we notice that the secondary structures labeled with the number 1 in Fig. \ref{Kottamia} and number 4 in Figure \ref{Wise} comprise the structures labeled as A1 and A2 (before the closest approach) and A3 and A4 (after the closest approach) in the 2011 FTN light curve \cite[see Fig. 1 in][]{Sickafoose2020}. In the same sense, structures labeled with numbers 2 and 5 in Figs. \ref{Kottamia} and \ref{Wise}, respectively, are compatible with structures A5 and A12 in Fig. 1 in \citet{Sickafoose2020}.

\begin{figure}[!ht]
    \centering
    \includegraphics[width=\hsize]{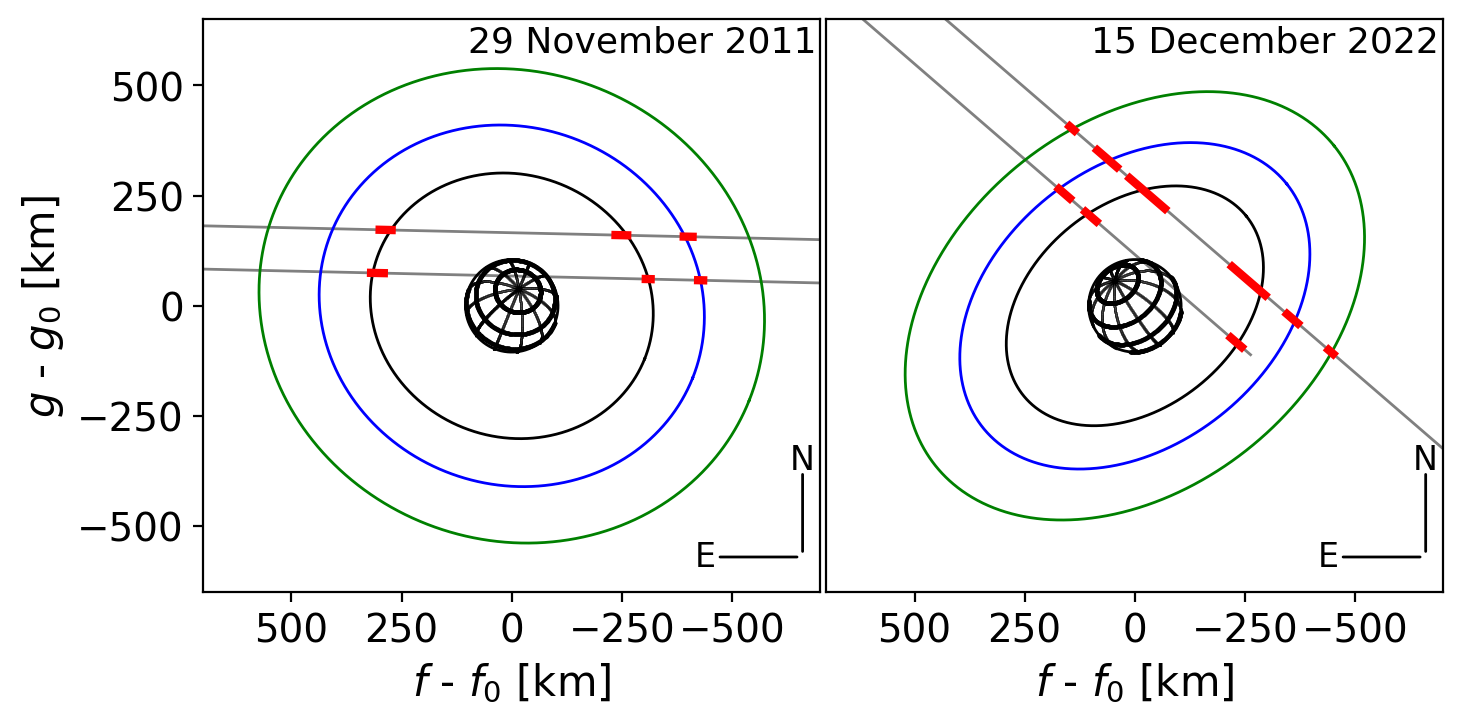}
    \caption{Sky plane plots of the structures that can fit the 2011 (left panel) and 2022 occultation features (right panel). The red segments correspond to the full extent of the extinction features. The grey straight lines correspond to the Kottamia and Wise chords.}
    \label{fig:joint_fit}
  \end{figure}

  Therefore, we can use these new detections of secondary structures from the 2022 event and the previous detections to improve the orbital parameters of these proposed rings. This was made using pipelines built with \textsc{sora} package \citep{GomesJr2022} by testing a range of pole orientations and radii for a circular ring centered in Chiron using a $\chi^2$ statistic \citep[more details in][]{Morgado2023,Pereira2023}, looking for the best fit to the innermost secondary structures since these are identified on both sides of the main body in all light curves and both events. The solution that best explains this detection is a circular ring with radius r = 325 $\pm$ 16 km with pole ecliptic coordinates $\lambda$ = 151$^\circ~\pm$ 8$^\circ$ and $\beta$ = 18$^\circ~\pm$ 11$^\circ$. If one uses the central moment of the long extinction features in the 2022 event instead of their full length, the ring radius is 315 $\pm$ 9 km, and the pole coordinates are $\lambda$ = 151$^\circ~\pm$ 3$^\circ$ and $\beta$ = 16$^\circ~\pm$ 4$^\circ$. 
  % \red \textbf{[CLP: Changing the approach: Fit the inner and outer rings separately, comparing the poles and checking if the solutions are compatible within the obtained uncertainties, thus reinforcing that the rings are co-planar] }
  Assuming that the adjacent structures are co-planar to this innermost ring-like structure, we calculate that the outermost structures lie at $\sim$442 km and $\sim$580 ~km away from the center of Chiron. The fits are shown in Fig. \ref{fig:joint_fit}.

 % \bla

\subsection{Discussion in connection with Chiron brightness enhancements}

The significant differences between the profiles of the 2011 occultation \citep[Fig. 2 of][]{Sickafoose2020} and those of our Fig. \ref{Kottamia} clearly show that Chiron cannot have a ring like that of Chariklo, which remains at a steady state. The structure in Chiron has evolved very significantly from 2011 to 2022. In the inserts of Fig. \ref{Kottamia}, we show that the ring model that explains the 2011 features, when convolved to the resolution achieved in 2022, cannot reproduce the observed curve in 2022. It reproduces the most profound drops but does not account for the rest of the extinction observed, meaning that there is more material around Chiron in 2022 than in 2011.

\begin{figure}
\centering
    \includegraphics[width=\columnwidth]{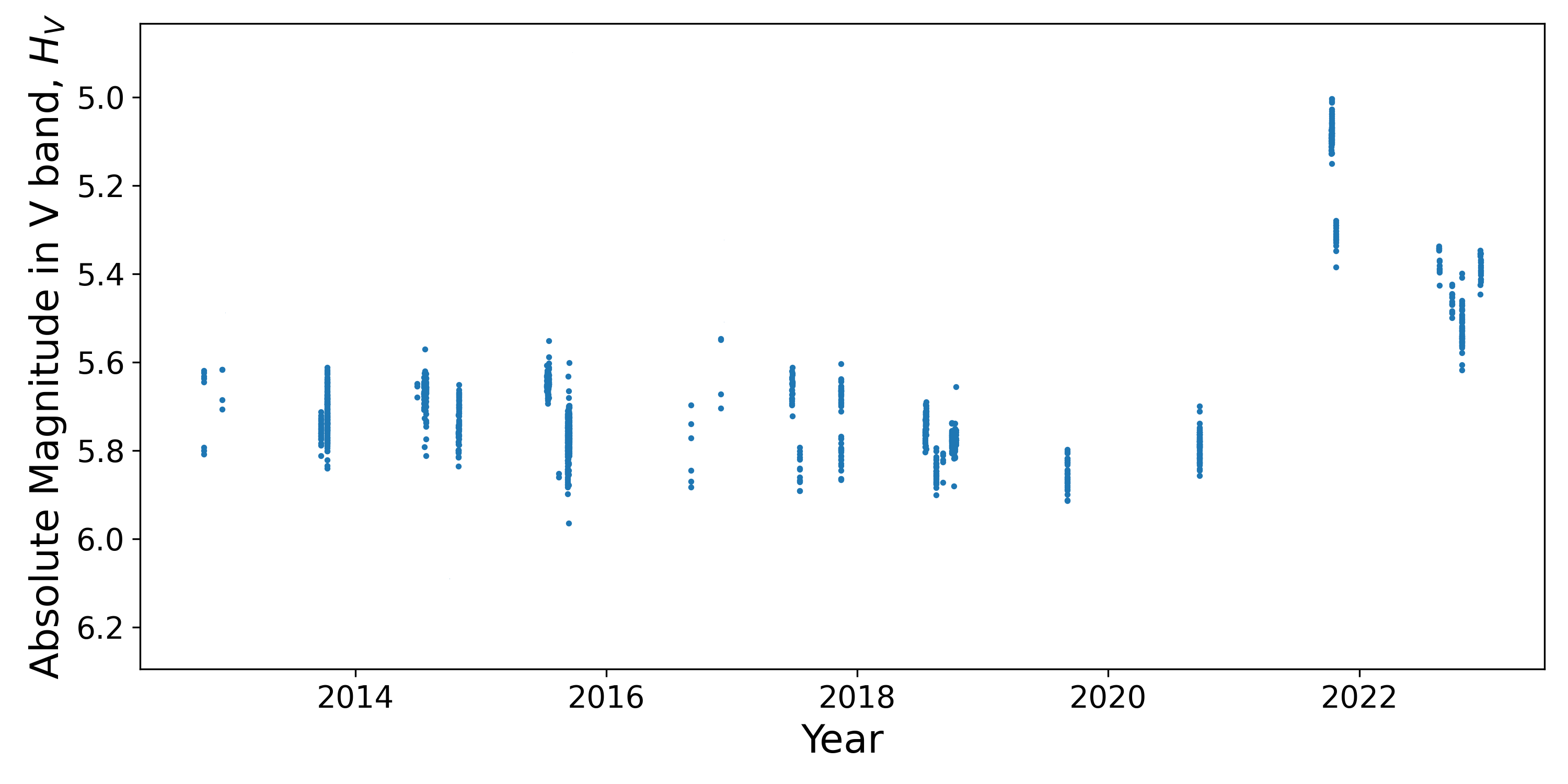}

  \caption{Absolute magnitude of Chiron in V band in the last ten years from the analysis of our imaging database using the procedures described in \cite{Morales2022}. A clear brightness increase by $\sim$0.6 mag is observed in 2021. The brightness of Chiron has not yet returned to the preoutburst level.}
  \label{AbsolutePhotometry}
\end{figure}

\cite{Dobson2021} reported a considerable brightness increase in Chiron sometime after February 8, 2021, and continued at least till June 18.
%(https://ui.adsabs.harvard.edu/abs/2021RNAAS...5..211D/abstract)
%Research Notes of the AAS, Volume 5, Issue 9, id.211.
%New or Increased Cometary Activity in (2060) 95P/Chiron. 
Our own photometry (corrected for heliocentric and geocentric distance as well as for phase effects using a phase coefficient of 0.15 mag/degree from our own linear fit to the reduced magnitude versus phase angle) also shows this remarkable brightness outburst within a data set spanning more than ten years. The increase was at least 0.6 mag. See Fig. \ref{AbsolutePhotometry}. Our photometry indicates that Chiron had not returned to its pre-outburst brightness at the occultation reported here, meaning that more dust/ice was present at the occultation than at its non-outburst state. \cite{Dobson2021} reported that the point spread function (PSF) of Chiron at outburst could not be distinguished from stellar PSFs, implying that the coma of Chiron or the dust structure causing the brightness increase was confined within a minimal angular diameter, consistent with our occultation observations, in which the most prominent feature extends below 1200 km in diameter: Such a size corresponds to an angular size that cannot be resolved with ground-based telescopes. Therefore, the material causing the brightness increase appears bound to the nucleus or escaping slowly.

%Spherical shells of material ejected from Chiron could cause symmetrical extinction features in occultations. However, the extinction caused by these potential shells would not vanish near the centrality, whereas the signal goes back to its normal level at the center of the occultation as seen in the Kottamia data set (and this is also the case in the Wise observatory data just prior to the occultation by the central body). This lack of extinction near the center of the body is better explained by an extended disk with enhancements in the form of ring-like structures and with a clearing near the central body due to the action of gravity, which removes all particles at least up to the 2:1 spin-orbit resonance \citep[see models by, e.g.][]{Sicardy2019}. Besides, the occultation features of both the 2011 and 2022 occultations can be reconciled with the ring parameters mentioned above, whereas shell structures would have different characteristics for 2011 and 2022.

%On the other hand, three clouds of dust or ice surrounding Chiron would give rise to more reflected light than what is observed given the measured optical depth in extinction, even using extremely low geometric albedos for the particles, whereas a ring structure on a plane would produce smaller amounts of reflected light. 

%I change the above discussion on shells in this short paragraph. We show here our preferred explanation. We do not have to rule out the shell structures at this point. Besides, there is no room for a short letter.
The material seems to be preferentially bound in a plane whose pole was derived in the previous section, with a clearing near the central body due to the action of gravity, which removes all particles at least up to the 2:1 spin-orbit resonance \citep[see models by e.g.][]{Sicardy2019}.

 It is interesting to point out that the brightness of Chiron in its 2021 maximum was almost identical in magnitude to the maximum brightness ever observed in Chiron, which corresponds to 1972 or 1973, as shown in Fig. 7 of \cite{Ortiz2015}, which comes from photometry compiled in \cite{Belskaya2010}. Moreover, it is curious that almost 50 years have elapsed between the two most significant brightness maxima. Note that 50 years is close to the orbital period of Chiron, which does not seem to be a mere coincidence. The geometric configuration at those dates might have resulted in an active area rich in CO ice or other volatiles illuminated by the Sun and generating a surge of activity. Another explanation could be that Chiron might cross the trajectory or the plane of a swarm of particles or debris from a disintegrated Centaur or comet whose impact shower might release dust or ice particles from the surface of Chiron or its rings. Note that the ecliptic crossing epochs for Chiron are 1976 and 2027 (descending nodes), not far from the epochs of brightness maxima. 

 %\subsection{A ring in the making?}

 %\blu Speculate here. \bla Chariklo's ring structure is steady and homogeneous, contrasting with what we see for Chiron. It is possible that a ring system is being formed at Chiron.

 \subsection{ A ring in the 3:1 resonance and structures outside the Roche limit?}

 %\subsection{Any ring in the 3:1 resonance? Implications for the density of the central body}

The ring of Haumea is located very close to its 3:1 spin-orbit resonance \citep{Ortiz2017} and this also appears to be the case for Chariklo, although the rotation period of Chariklo is not as clearly determined as in the case of Haumea, and its central mass is not known. For Quaoar, the main ring is close to the 3:1 resonance \citep{Morgado2023,Pereira2023}. Therefore, this appears to be a common feature and may cause the confinement of the ring structures in the outer solar system bodies that show non-axisymmetric gravity potentials \citep{Sicardy2021}.

If the same behavior happens with the densest structure in Chiron, one can determine the body's mass and try to get a density for it, which is a very relevant parameter to understanding the basic physics of the TNOs and Centaurs. Unfortunately, we lack a good three-dimensional shape for Chiron, so its volume cannot be accurately computed, and thus no accurate density can be derived. To give an idea, assuming a plausible volume-equivalent diameter of 210 km, 
consistent with the volume-equivalent diameter of 196 $\pm$ 34 km from the three-dimensional shape recently constrained by \citep{BragaRibas2023},
the density would be 933 kg/m$^3$. This is at the high end of the density estimates in \cite{Bierson2019} for bodies of Chiron's size.
% If the draft paper by Felipe on the previous occultations is submitted, we can use his ellipse fit to one of the events to update the size and compute a better density, but he may be left for the future.

At least the particle concentrations beyond 423 km would seem to be outside the Roche limit of the main body. % and should not remain without accreting for a long time unless we are in a similar situation as in Quaoar because the collision of the ring particles is more elastic than thought before or for other unknown reasons. 
 % Following the same approach as in \cite{morgado2023}, we can calculate the density the material in orbit should have to avoid accretion. Assuming a density of 800 kg/m3 for Chiron's (the expected density for a TNO of its size) and an equivalent diameter of 210 km, we can determine its mass and use the same equation as in \cite{morgado2023}. It turns out that for the structure at 423 km, the density of the material would have to be 100 kg/m3 not to accrete, which is significantly smaller than the typical value of 400 kg/m3, as discussed in \cite{morgado2023}. Hence we have a similar situation with Chiron as with Quaoar, but we are using a guess density for Chiron. Hence, a definite conclusion cannot be made.
 %Assuming a conservatively large radius of 120 km for Chiron, the ring structure at 423 km is at least at 3.5 Chiron radii, far away from the classical Roche limit given by: $2.4 R (\rho_p/\rho_s)^{1/3} $ where R is the radius of the primary and $\rho_p$ and $\rho_s$ are the densities of the primary and the material in orbit respectively. Even using an extremely high density for Chiron of 1400 kg/m$^3$ and a very low density for the ring material of 400 kg/m$^3$, the radial distance to Chiron would not be larger than 3.5 radii. Hence, at least one of the ring structures would be outside the Roche limit, similar to what has been recently observed for Quaoar \citep{Morgado2023}.
As done in \cite{Morgado2023} to show that the ring of Quaoar is outside the Roche limit, we can use the formula that relates the distance from the body with the critical Roche density that a satellite would have in order not to disrupt 
$a_{Roche} = [3 M_C/(\gamma \rho)]^{1/3}$ where $\rho$ is the critical density $M_C$ is the central mass, $a_{Roche}$ is the distance from the body and $\gamma= 1.6$ as in \cite{Morgado2023}. 
If we use Chiron's mass derived from the 3:1 resonance constraint, the Roche critical density would be 271 kg/m$^3$, which is well below the threshold value of 400 kg/m$^3$, so the structure at 423 km would be outside the Roche limit. If instead of using a mass for Chiron from the 3:1 resonance constraint, we derive a range of masses using a reasonable density interval of 600 to 900 kg/m$^3$ taken from \cite{Bierson2019} for Chiron's approximate effective diameter, we can derive other estimates of the Roche critical density. These would range from 120 to 180 kg/m$^3$, again, well below the 400 kg/m$^3$ threshold adopted in \cite{Morgado2023}, implying that the ring would be outside the Roche limit.

\section{Conclusions}
Through a stellar occultation, we have found that Chiron is currently surrounded by a tenuous disk of $\sim$ 580 km in radius, with dense ring or ring-like concentrations at 325$\pm$16 and 423$\pm$11 km, whose pole has ecliptic coordinates $\lambda=151^\circ \pm 8^\circ$ and $\beta=18^\circ \pm 11^\circ$. 
%The innermost ring is consistent in diameter and pole with secondary occultation features observed in previous events reported in the literature. 
%The other extinction features are reported here for the first time and might be related to the brightness outburst that Chiron experienced in 2021. 
Some of these features are consistent with secondary occultation extinction features observed in previous events reported in the literature. However, they are considerably enhanced in 2022, possibly due to the brightness outburst that Chiron experienced in 2021, which has not receded entirely yet.
%The extinction features observed in 2022 appear to be of a larger extent and maybe depth than in 2011.
At least the outermost structure seems to be outside the Roche limit, as is the case for Quaoar's rings. It is unclear whether the phenomenon that caused the observed brightness outburst of $\sim$0.6 mag in 2021 can feed the rings with material or whether some phenomena in the rings might have caused the brightness outburst. The brightness increase left Chiron at a high brightness level, near the maximum brightness of Chiron in the early seventies, around 50 years ago, near its orbital period. 
% Quasi-periodic collisional or sublimation scenarios might be at play.
%--------------------------------------------------- One column table
% \begin{table}
% \caption[]{Occulted star information. JHK magnitudes from 2MASS.}
% \label{KapSou}
 % $$ 
 % \begin{array}{p{0.5\linewidth}l}
   % \hline
   % \noalign{\smallskip}
   % RA(J2000) & Dec(J2000) \\
   % \noalign{\smallskip}
 % \hline
 % \noalign{\smallskip}
 % Gaia ID & 2556932574069626624 \\
 % RA(J2000) & 00h39m58.7191s \\
 % Dec(J2000) & +06^\circ 16' 08.858'' \\
 % G magnitude & 12.7 \\
 % J magnitude & 12.1 \\
 % H magnitude & 12.0 \\
  % K magnitude & 12.0 \\
    % Cox \& Stewart 1969 & 5000 \leq \\
 % \noalign{\smallskip}
  % \hline
 % \end{array}
 % $$ 
 % \end{table}

%--------------------------------------------------- One column table
% \begin{table}
% \caption[]{Opacity sources.}
% \label{KapSou}
% $$ 
% \begin{array}{p{0.5\linewidth}l}
% \hline
% \noalign{\smallskip}
% Source & T / {[\mathrm{K}]} \\
% \noalign{\smallskip}
% \hline
% \noalign{\smallskip}
% Yorke 1979, Yorke 1980a & \leq 1700^{\mathrm{a}} \\
%% Yorke 1979, Yorke 1980a & \leq 1700 \\
% Kr\"ugel 1971 & 1700 \leq T \leq 5000 \\
% Cox \& Stewart 1969 & 5000 \leq \\
% \noalign{\smallskip}
% \hline
% \end{array}
% $$ 
% \end{table}

% One column figure
%----------------------------------------------------------------- 
% \begin{figure}
% \centering
% %%%\includegraphics[width=3cm]{empty.eps}
% \caption{Vibrational stability equation of state
% $S_{\mathrm{vib}}(\lg e, \lg \rho)$.
% $>0$ means vibrational stability.
% }
% \label{FigVibStab}
% \end{figure}
%-----------------------------------------------------------------

\begin{acknowledgements}
   Part of this work was supported by the 
Spanish projects PID2020-112789GB-I00 from AEI and Proyecto
de Excelencia de la Junta de Andalucía PY20-01309. Financial support from the grant CEX2021-001131-S funded by MCIN/AEI/ 10.13039/501100011033 is also acknowledged. This research is partly based on observations taken with the 1.88-m telescope at the Kottamia Astronomical Observatory (KAO), operated by researchers at the National Research Institute of Astronomy and Geophysics (NRIAG), Egypt. The Egyptian team acknowledges support from Science, Technology \& Innovation Funding Authority (STDF) under grant number 45779.
C.L.P is thankful for the support of the CAPES and FAPERJ/DSC-10 (E26/204.141/2022).
\end{acknowledgements}

% \bibliographystyle{aa} % style aa.bst
% \bibliography{Yourfile} % your references Yourfile.bib
\bibliographystyle{aa}
\bibliography{references.bib}

% Appendix A
\appendix
\section{Observational circumstances}

\begin{table}[!ht]
  \centering
  \caption{Observing details}
  \begin{tabular}{c c c c c c c c} 
  \toprule
  \toprule
  \multirow{3}{*}{Site} & Latitude [$^{\circ}$ ' ''] & Aperture [mm] & Exp. Time & & \\ 
              & Longitude [$^{\circ}$ ' ''] & Detector & Cycle & Observers & Detection\\ 
              & Altitude [m] & Filter & [s] & & \\ 
  \midrule
    Kottamia & 29 56 02.4 & 1,880 & 3.0 & \multirow{2}{*}{A. Takey} & \\
  Astronomical Obs. & 31 49 37.2 & E2V 42-40 2k & 4.5 & \multirow{2}{*}{A. Fouad} & Structures \\ 
  Cairo, Egypt & 476 & SDSS-g' & & & \\ 
  \midrule
    Wise Observatory & 30 35 48.59 & 457.2 & 3.0 & & \multirow{2}{*}{Nucleus +} \\ 
  Mitzpe Ramon & 34 45 44.14 & QSI683 & 7.5 & S. Kaspi & \multirow{2}{*}{Structures} \\ 
  Israel & 862.3 & Clear & & & \\ 
  \midrule
  TUBITAK & 36 49 17.07 & 1,000 & 0.5 & \multirow{2}{*}{Y. Kilic, O. Erece} & \\
  National Obs. & 30 20 07.98 & SI 1100 & 5.1 & \multirow{2}{*}{D. Koseoglu, E. Ege} & Negative \\ 
  Antalya, Turkey & 2,538.7 & Clear & & & \\ 
  \midrule
  \hline
  \end{tabular}
  \label{tab:observ_circums}
\end{table}

\end{document}